\documentclass[prb, twocolumn,showpacs,superscriptaddress, preprintnumbers,amsmath,amssymb]{revtex4}

\usepackage{graphicx, color}
\begin{document}


\title{Fermi surfaces, electron-hole asymmetry and correlation kink in a three-dimensional Fermi liquid LaNiO$_3$}




\author{R.~Eguchi}
\altaffiliation[E-mail: ]{ritsuko@spring8.or.jp}
\affiliation{RIKEN SPring-8 Center, Sayo-cho, Sayo-gun, Hyogo 679-5148, Japan}
\affiliation{Institute for Solid State Physics, University of Tokyo, Kashiwanoha, Kashiwa, Chiba 277-8581, Japan}

\author{A.~Chainani}
\author{M.~Taguchi}
\author{M.~Matsunami}
\author{Y.~Ishida}
\author{K.~Horiba}
\affiliation{RIKEN SPring-8 Center, Sayo-cho, Sayo-gun, Hyogo 679-5148, Japan}

\author{Y. Senba}
\author{H. Ohashi}
\affiliation{JASRI/SPring-8, Sayo-cho, Sayo-gun, Hyogo 679-5198, Japan}

\author{S.~Shin}
\affiliation{RIKEN SPring-8 Center, Sayo-cho, Sayo-gun, Hyogo 679-5148, Japan}
\affiliation{Institute for Solid State Physics, University of Tokyo, Kashiwanoha, Kashiwa, Chiba 277-8581, Japan}


\date{\today}

\begin{abstract}
We report the three-dimensional (3-D) momentum-resolved soft x-ray photoemission spectroscopy of the Fermi liquid LaNiO$_3$. The out-of-plane and in-plane cuts of the 3-D electron- and hole-Fermi surfaces (FSs) are observed by energy- and angle- dependent photoemission measurements. The energy bands forming the electron FS suggest an $\omega^2$ dependence of the imaginary part of the self-energy and a `correlation kink' at an energy scale of 0.25 eV. In contrast, the bands which form nesting character hole FSs do not show kinks and match local density approximation calculations. The results indicate a momentum-dependent mass renormalization, leading to electron-hole asymmetry in strongly correlated LaNiO$_3$.           
\end{abstract}


\maketitle

\section{INTRODUCTION}

Transition metal perovskite oxides provide the widest range of correlation derived phenomena such as superconductivity, metal-insulator transitions (MITs), magnetic ordering, colossal magneto resistance, etc. \cite{Imada} While momentum-resolved studies of the two dimensional high-$T_c$ cuprates and other low dimensional correlated materials have revealed novel aspects of their electronic structure and Fermi surfaces (FSs), \cite{Shen} there is no report of a FS study of a three-dimensional (3-D) correlated oxide. Recent studies have established that oxide films with three-dimensional electronic structures are fertile ground for discovering novel properties for device applications, such as interfacial metallicity, carrier-controlled transparent ferromagnetism, giant thermoelectricity, etc. \cite{Ohtomo,Philip,Ohta} These engineered properties arise from correlated electrons, as evidenced from momentum($k$)-averaged experiments of transport, magnetism, thermodynamics, etc. However, there is no momentum-resolved study to date showing an $\omega^2$ dependence of the imaginary part of the self-energy for a 3-D Fermi liquid. \cite{Imada,Shen,Landau} Among the perovskite oxides, the series $R$NiO$_3$ ($R$: rare earth) is well known to show interesting behavior such as MITs, spin- and charge-ordering. LaNiO$_3$ with a trivalent Ni$^{3+}$ ($t_{2g}^6e_g^1$) is the reference material of the series, crystallizing in a 3-D nearly cubic structure with a small rhombohedral distortion ($\beta = 90.41 ^{\circ}$). It exhibits enhanced Pauli paramagnetic metal behavior down to low temperature ($T$). LaNiO$_3$ undergoes an oxygen vacancy controlled correlated metal to antiferromagnetic insulator transition, while the other members (e.g., $R$ = Pr, Nd, Sm) show MIT as a function of $T$. \cite{Munoz,Zhou}  
The $T$ dependent MIT coupled antiferromagnetic transition results in a ($h$/2 0 $l$/2) reflection due to charge-order (more precisely, charge disproportionation of the type Ni$^{3+\delta}$-Ni$^{3-\delta}$), and not due to orbital ordering, as was recently characterized by resonant soft and hard x-ray scattering. \cite{LorenzoScagnoli} Interestingly, the magnetic susceptibility of the entire series in the metallic phase is that of an enhanced Pauli metal due to the NiO$_{6}$ matrix, and far from a Stoner type ferromagnetic instability. \cite{Zhou,Sreedhar} The specific heat and susceptibility data, as well as $T^2$ dependence of resistivity of LaNiO$_3$ indicate that this compound is well described as a strongly correlated system close to a MIT. \cite{Sreedhar,Xu,Rajeev} An enhanced effective mass ($m^*$ $\sim$ 10$m_0$) is well established from thermopower and specific heat measurements. \cite{Rajeev,Xu} Thus, LaNiO$_3$ is a genuine 3-D Fermi liquid with correlated 3$d$ electrons in a quarter filled $e_g$ band.

Bulk single crystal growth of LaNiO$_3$ has not been possible to date. However single crystalline epitaxial thin films have been successfully synthesized for device applications and provide a very suitable metallic template for electrical contacts to perovskite films. \cite{ChenDobin} In a recent study, we succeeded in fabricating single crystalline epitaxial thin films and studying it by angle-integrated photoemission spectroscopy (PES). \cite{Horiba1} We reported the observation of a clear sharp $e_g$ derived peak at the Fermi level ($E_{\rm{F}}$), which could not be detected in earlier studies. \cite{Barman} This peak gets gapped across the oxygen vacancy controlled MIT in LaNiO$_{3-x}$. A narrow band at $E_{\rm{F}}$ is typical of correlated metals, but momentum or angle-resolved PES is necessary to directly probe 3-D FSs and renormalization of band dispersions. Such studies require an accurate determination of the in-plane and out of plane components of the momentum vector of an electron, and is an intrinsically difficult experiment for 3-D systems requiring tunable photons from a synchrotron source. Recent studies have reported soft x-ray angle-resolved photoemission spectroscopy (ARPES) of elemental metals \cite{HussainChKamakura} and correlated-electron systems. \cite{VenturiniYanoFujimori} The role of non-direct transitions could be identified at high temperature in elemental metals. For {\it f}-electron systems, dispersive bands and even FS crossings have been identified. However, the momentum dependence of renormalized band dispersions due to electron-electron correlations has not been addressed. Here, we report the momentum-resolved PES of LaNiO$_3$ using soft x-rays. The aim of the present work is to determine the FSs and the role of correlations in a 3-D Fermi liquid.

Recent theory showed that kinks can appear in the band dispersion of 3-D materials as a consequence of correlations, based on dynamical mean field theory in combination with the local density approximation (LDA+DMFT) calculations. \cite{NekrasovByczuk} These calculations carried out for the case of an energy dependent, but momentum independent self-energy, suggested consistency with low energy ARPES of SrVO$_3$. \cite{Yoshida} Kinks are often found in systems with strong electron-phonon coupling within 100 meV below $E_{\rm{F}}$, for example, as in high-$T_c$ superconductors. \cite{Lanzara} Moreover kinks on a higher energy scale, about 300-500 meV, were also reported in high-$T_c$ superconductors. \cite{Valla,Meevasana} These anomalies suggest that kinks can occur due to a variety of reasons such as electron-phonon interaction, short-range Coulomb interaction, etc. Since LaNiO$_3$ is a strongly correlated metal, one may expect to see anomalies in its momentum-resolved electronic structure.

\section{EXPERIMENTAL DETAILS}

The LaNiO$_3$ thin films were grown on SrTiO$_3$ single-crystal substrates.  A sintered stoichiometric LaNiO$_3$ pellet was used as an ablation target.  A Nb-doped yttrium aluminum garnet laser was used in its frequency-tripled mode ($\lambda$ = 355 nm) at a repetition rate of 1 Hz.  The SrTiO$_3$ substrates were annealed at 900 $^{\circ}$C at an oxygen pressure of 1 $\times$ 10$^{-4}$ Pa before deposition.  The substrate temperature was set to 650 $^{\circ}$C and the oxygen pressure was 10 Pa during the deposition.  The LaNiO$_3$ films were subsequently annealed at 400 $^{\circ}$C for 30 minutes at atmospheric pressure of oxygen to remove vacancies. After cooling the sample to below 100 $^{\circ}$C and evacuating the growth chamber, the surface morphology and crystallinity of the fabricated LaNiO$_3$ films were checked by an in situ observation of reflection high-energy electron diffraction (RHEED) patterns. Further details on the characterization are described in Ref.\ \onlinecite{Horiba1}.

Soft x-ray ARPES experiments were carried out using a high-resolution synchrotron radiation PES system with a Gammadata-Scienta \linebreak SES2002 spectrometer combined with a pulse laser deposition chamber at undulator beamline BL17SU, SPring-8. \cite{Horiba2Ohashi} The ARPES measurements were carried out in a vacuum of 5 $\times$ 10$^{-8}$ Pa, at 30 K to minimize non-direct transition effects. The energy resolution was set to $\sim$150 and 200 meV and the angular resolution was $\pm$ 0.2 degree, which corresponds to a momentum resolution of $\pm$ 0.043 {\AA}$^{-1}$ at $h\nu$ = 630 eV

The high symmetry momentum points ($\Gamma$ and X) and the Fermi crossings were accurately determined from energy dependent normal emission spectra and a free electron final state, taking into account the photon momentum into consideration. Thus,
\[k_{z} = \sqrt {\frac{2m}{{\hbar^2 }} \left( E_{\rm kin}cos^{2}\theta + V_{0} \right)} -k_{\perp h\nu} ,\]
where $k_{z}$ is the normal component of the momentum, $m$ is the electron mass, $E_{\rm kin}$ is the kinetic energy of the electrons,  $k_{\perp h\nu}$ is the component of the photon momentum perpendicular to the surface and $\theta$ is the emission angle. An inner potential of $V_{0}$ = 10 eV was assumed to obtain consistent results. We required a photon momentum correction of $\sim$14 $\%$ and $\sim$15 $\%$ of 2$\pi$/c at 630 eV ($\Gamma$-X-$\Gamma$ cut) and 710 eV (X-M-X cut), respectively, where c is the pseudo cubic lattice parameter of LaNiO$_{3}$. The high-symmetry points were further calibrated by angle dependent in-plane measurements at fixed photon energies.  All the intensity maps shown in the paper are measured data over the displayed ranges and no symmetrization has been used to obtain the maps.


\begin{figure}
\begin{center}
\includegraphics[scale=1]{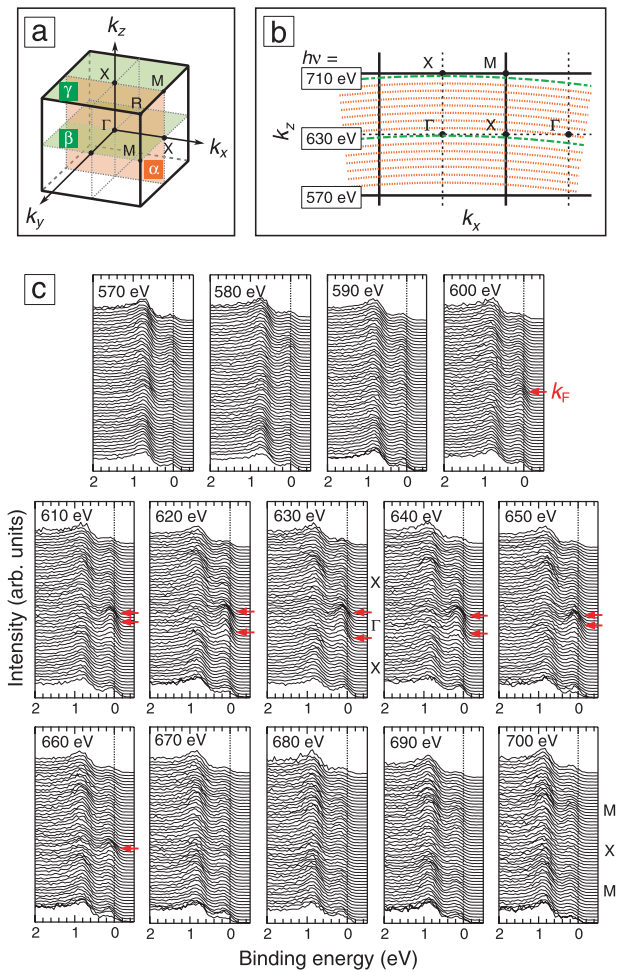}
\end{center}
\caption{(Color online)  (a) The schematic of a cubic Brillouin zone. (b) The momentum space region of the ARPES measurements. (c) The EDCs at photon energies between $h\nu$ = 570 and 700 eV at normal emission.
}\end{figure}

\section{RESULTS AND DISCUSSION}

Figures 1(a) and 1(b) show the schematic of a cubic Brillouin zone and the region in momentum space of the ARPES measurements reported here (15 cuts, using $h\nu$ = 570-710 eV, steps of 10 eV). While bulk LaNiO$_3$ as well as thin film LaNiO$_3$ show a small rhombohedral distortion and are not simple cubic, the band structure of rhombohedral LaNiO$_3$ is well explained by zone folding of a cubic Brillouin zone into a rhombohedral Brillouin zone. \cite{Hamada} Hence, we adopt a pseudo cubic notation for discussing the PES of LaNiO$_3$. Figure 1(c) shows the energy distribution curves (EDCs) at photon energies between $h\nu$ = 570 and 700 eV at normal emission. EDCs at all photon energies show intense features with small dispersion around 0.8 eV, corresponding to the Ni $t_{2g}$ bands. In addition, EDCs for 600-660 eV clearly show a dispersive band crossing at the $E_{\rm{F}}$ (red arrows) around the $\Gamma$ point. This dispersive band originates in Ni $e_{g}$ states and forms a small electron pocket as predicted by the local density approximation (LDA) band calculation. The band disappears for energies below 600 eV and above 660 eV-photon energy and an increase in intensity is observed close to the M point at $h\nu$ = 570 and 700 eV.

Figure 2(a) shows the FS mapping in a vertical ($k_z$-$k_x$) plane `$\alpha$' of Brillouin zone as shown in Fig.\ 1(a), which is obtained by a plot of the integrated intensity from -0.05 to 0.05 eV-binding energy in EDCs. A small circle centered at $\Gamma$ point is observed, while no intensity is observed around X point. The existence of a nearly spherical small FS centered at $\Gamma$ point, corresponding to the electron FS, was predicted by band calculations. \cite{Hamada} 
From photon energy($k_z$)-dependent ARPES measurements, we can decide the photon energy tracing the $\Gamma$-X and X-M directions as shown in Fig.\ 1(b). In order to observe the in-plane cuts of the FSs, we measured ARPES at a fixed photon energy of 630 eV ($k_z$ $\sim$ 0) and 710 eV ($k_z$ $\sim$ 0.5). The FS mapping in horizontal ($k_x$-$k_y$) planes `$\beta$' and `$\gamma$' of Brillouin zone in Fig.\ 1(a) were obtained as shown in Figs.\ 2(b) and 2(c). In `$\beta$' plane, the small electron FS around $\Gamma$ point is again observed. This result indicates that the 3-D sphere-like FS around $\Gamma$ point can be observed by ARPES experimentally. The square-like intense area around M points, obtained from raw data without symmetrization, corresponds to projection of the hole FS centered at R point. In `$\gamma$' plane, no small FS is observed. Instead large FSs centered at R point were observed. From the complete data set of energy and angle dependent FS maps, the experimental FS obtained by soft x-ray ARPES are in overall agreement with that predicted by the band calculation. \cite{Hamada} However, the actual band dispersions reveal an important difference in electron- and hole-FSs as discussed in the following.

\begin{figure}
\begin{center}
\includegraphics[scale=1]{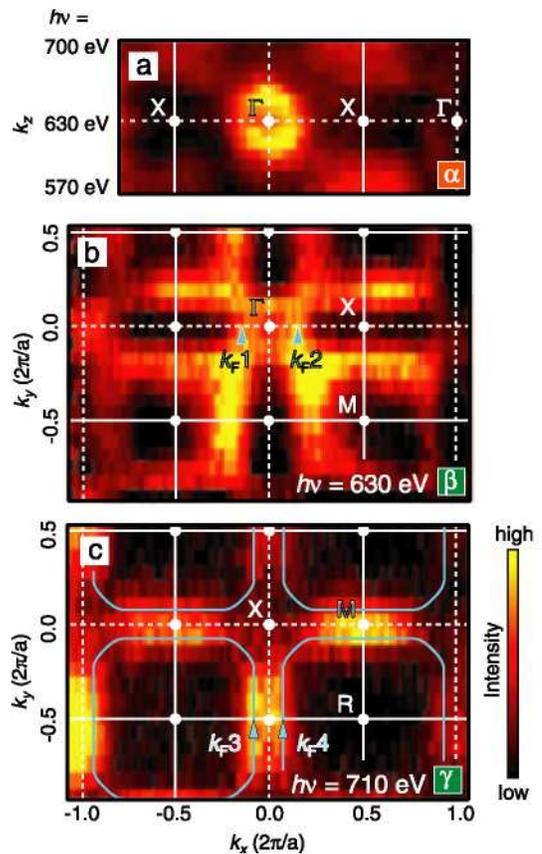}
\end{center}
\caption{(Color online)  
Fermi surface mapping in (a) a vertical ($k_z$-$k_x$) plane `$\alpha$', and (b) horizontal ($k_x$-$k_y$) planes `$\beta$' and (c) `$\gamma$' of Brillouin zone in Fig.\ 1(a). Solid white lines correspond to the cubic Brillouin zone and dotted white lines correspond to the high symmetry lines. $k_{\rm{F}}$1-$k_{\rm{F}}$4 indicate the MDC peak positions at $E_{\rm{F}}$ in Figs.\ 3(c) and 3(d). Blue lines show nesting character hole FSs. All the intensity maps are measured data over the displayed range and no symmetrization has been used to obtain the maps. 
}\end{figure}

\begin{figure}
\begin{center}
\includegraphics[scale=1]{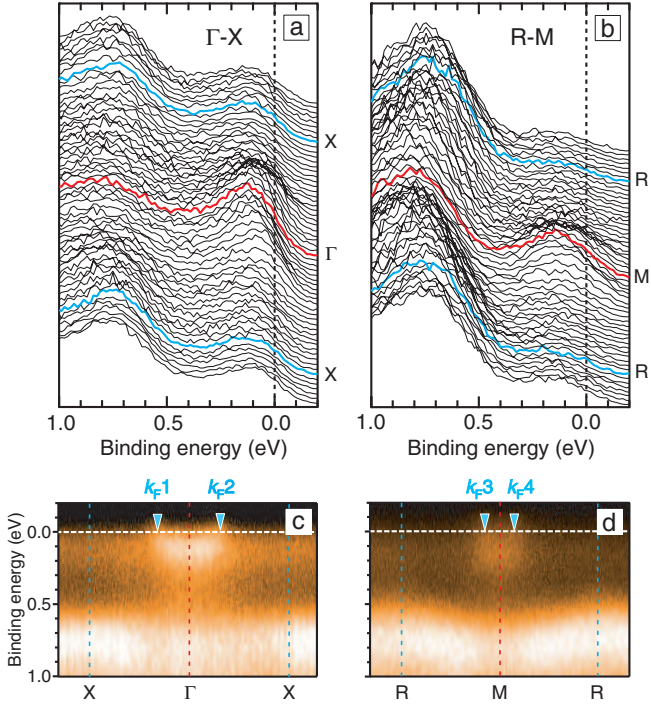}
\end{center}
\caption{(Color online)  
(a) and (b) EDCs along $\Gamma$-X and R-M direction, respectively. (c) and (d) The intensity maps of EDCs. $k_{\rm{F}}$1-$k_{\rm{F}}$4 indicate the MDC peak positions at $E_{\rm{F}}$. 
}\end{figure}

\begin{figure}
\begin{center}
\includegraphics[scale=1]{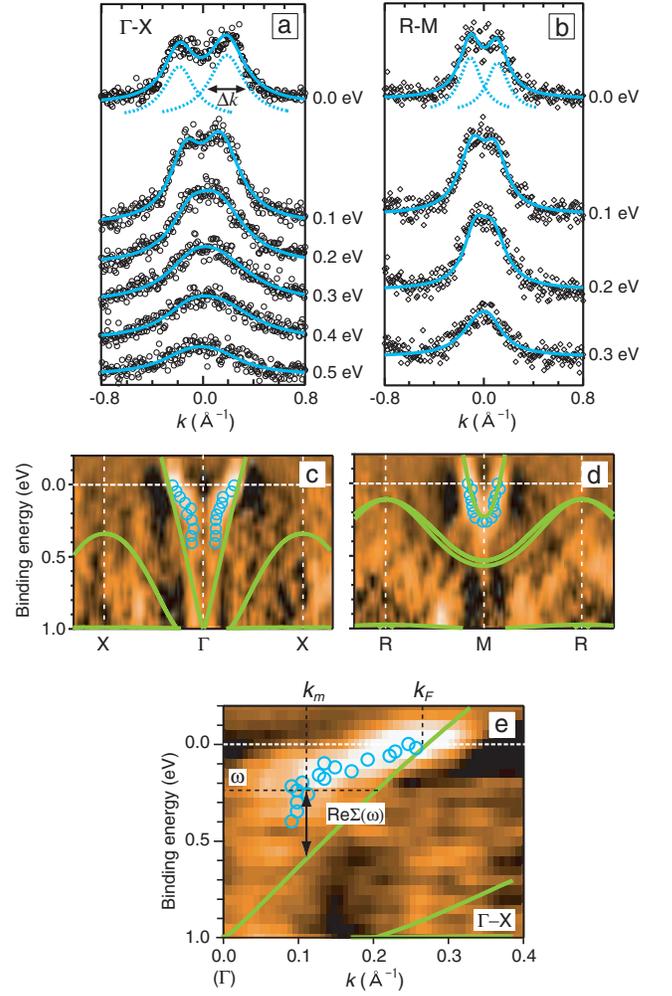}
\end{center}
\caption{(Color online)  
(a) and (b) MDCs with fitted spectra using two Lorentzians. (c) and (d) The intensity map of the second derivative MDCs with LDA band calculation (green lines). The experimental band dispersion derived from the peak positions of second derivative MDCs is shown as the blue circles. (a) and (c) were measured along $\Gamma$-X direction, while (b) and (d) were measured along R-M direction. (e) Expanded view of the intensity map of the second derivative MDCs with LDA band calculation (green lines) in $\Gamma$-X direction.
 }\end{figure}

\begin{figure}
\begin{center}
\includegraphics[scale=1]{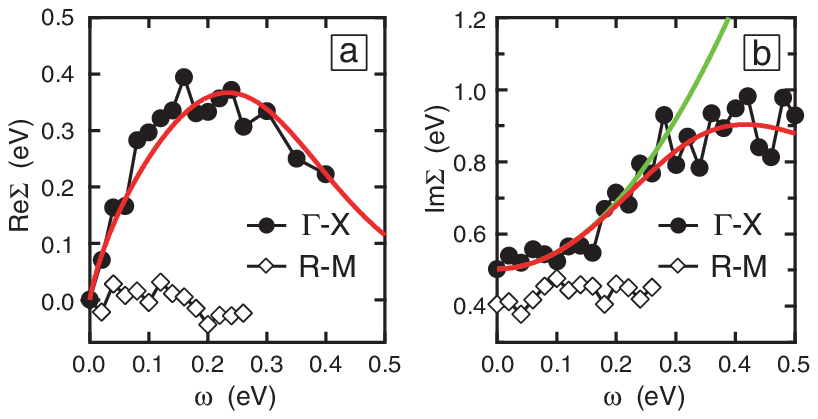}
\end{center}
\caption{(Color online)  
(a) Re$\Sigma$($\omega$) and (b) Im$\Sigma$($\omega$) obtained from experiment (dots) and a Kramers-Kronig analysis (red lines). Green line indicates Im$\Sigma$($\omega$) for a Fermi liquid. 
}\end{figure}

In order to discuss the band structures forming these FSs, we measured the ARPES in the high symmetry lines with detailed momentum steps and an energy resolution $\Delta{E}$ $\sim$ 150 meV. The EDCs  [Figs.\ 3(a) and 3(b)] and the intensity plots [Figs.\ 3(c) and 3(d)] along $\Gamma$-X and R-M directions are shown in Fig.\ 3. The FS crossing $k_{\rm{F}}$ points are labeled as $k_{\rm{F}}$1-$k_{\rm{F}}$4. The intensity plots in both directions [Figs.\ 3(c) and 3(d)] show a intense feature around 0.0-0.2 eV at $\Gamma$ and M points, corresponding to the electron band and the hole band derived from Ni 3$d$ $e_g$ states, respectively. While the band bottom at M point is around 0.25 eV, the band bottom at $\Gamma$ point is not at 0.25 eV. The intensity remains in high binding energy region (about 0.5-1.0 eV) at $\Gamma$ point [Fig.\ 3(c)], indicating that the bands extend to high binding energies. Because the $t_{2g}$ bands also appear above 0.5 eV at X point, the band bottom is not clear. This behavior contrasts with the hole band at M point [Fig.\ 3(d)], in which there is no intensity around 0.3-0.5 eV at M point. Therefore, the electron band at the $\Gamma$ point extends to at least 0.5 eV-binding energy.

Figures 4(a) and 4(b) show the momentum distribution curves (MDCs) plotted in steps of 0.1-eV binding energy. We fitted the MDCs to two Lorentzians to estimate the MDC peak width. For the $\Gamma$ centered electron FS, the mean free path of carriers is obtained as $l$=1/$\Delta{k}$ $\sim$ 3 {\AA} from the MDC width at $E_{\rm{F}}$, consistent with transport. \cite{Rajeev}  The mean free path of the carriers around $\Gamma$ point are smaller than that around M point.  
Moreover, to emphasize the band dispersion, we apply the second derivative for the MDCs and plot it as intensity map in Figs.\ 4(c) and 4(d). Blue circles indicate the peak positions of second derivative MDCs, corresponding to the experimental band dispersion. The theoretical band calculation result is also plotted on the experimental intensity plot. The theoretical calculation was done using the local density approximation (LDA) based full-potential linearized augmented plane-wave method \cite{wien2k} in the cubic symmetry with the lattice constant of a = 5.45/$\sqrt{2}$ = 3.85 {\AA} for the LaNiO$_3$ bulk crystal. \cite{Munoz} From the comparison between the experimental and the calculated band dispersion, the band forming the electron FSs shows clear differences from the calculated band, while the band forming the hole FSs almost matches the calculated band dispersion. A clear kink structure at $\sim$0.25 eV in the band forming the electron FS is observed in Fig.\ 4(e).

We have determined the real and imaginary parts of the self-energy [$\Sigma(\omega)$] from ARPES spectra, by neglecting its $k$-dependence and approximating the bare band dispersion to be linear, as is the usual procedure. \cite{Valla} The Re$\Sigma(\omega)$ and Im$\Sigma(\omega)$ are derived  from the widths  $\Delta{k}$ and peak positions $k_m$ of the MDC peaks, and the bare velocity $v_0$ as shown in the following equation, 
\[ k_m = k_F + [\omega - {\rm Re}\Sigma(\omega)]/v_0, \]
\[\Delta{k} = 2{\rm Im}\Sigma(\omega)/v_0 . \]   

The results of such an analysis are shown in Figs.\ 5(a) and 5(b).
The Im$\Sigma(\omega)$ at $E_{\rm{F}}$ [Im$\Sigma(0)$ $\sim$ 0.5 eV] is rather high, along $\Gamma$-X. In contrast, the Re$\Sigma(\omega)$ at $E_{\rm{F}}$ is 0 along $\Gamma$-X, since the band crossing in experiment matches the band structure calculations. The high Im$\Sigma(\omega)$ at $E_{\rm{F}}$ along $\Gamma$-X is assumed to be a constant in terms of an impurity scattering term. Although the Im$\Sigma(0)$ is large, a similar situation has been reported with Im$\Sigma(0)$ $\sim$ 0.25 eV for under-doped cuprates. \cite{Valla}   
Using an analytic function $\propto (a\omega)^2/[1+(a\omega)^4]$ to model the Im$\Sigma$($\omega$) of the Fermi liquid, we have checked  the self-consistency between the Re$\Sigma(\omega)$ and Im$\Sigma(\omega)$ using the Kramers-Kronig relations [red lines in Figs.\ 5(a) and 5(b)]. The behavior of Re$\Sigma(\omega)$ and Im$\Sigma(\omega)$ is remarkably similar to the high-$T_c$ superconductors having the high-energy kink. \cite{Valla,Meevasana} The deviation from $\omega^2$ behavior [green line in Fig.\ 5(b)] identifies the kink at 0.25 eV. The present results suggest a Fermi liquid behavior for LaNiO$_3$, consistent with transport studies. \cite{Sreedhar,Xu,Rajeev} The energy scale of  0.25 eV is very large and cannot be associated with a phonon energy scale in LaNiO$_3$. The kink behavior leads to a picture as was shown by LDA+DMFT calculations for strongly correlated electron systems. \cite{NekrasovByczuk} On the other hand, the hole bands do not show a renormalization kink like the electron bands. This result indicates that the hole carriers do not have an enhanced effective mass like the electron carriers. The difference of the mean free path between the electron and hole carriers is shown in the MDC width at $E_{\rm{F}}$. LaNiO$_3$ is discussed in terms of a $t_{2g}^{6}e_g^{1}$ state and shows $n$-type conductivity in Hall measurements. \cite{Noun} In electron- and hole-doped $R_{1-x}A_{x}$NiO$_3$, it was found that a significant electron-hole asymmetry leads to the suppression of the MIT temperature by the extra carriers. \cite{Munoz2}

The mass renormalization factor is estimated as $m^*$/$m_b$ $\times$ $m_b$/$m_0$ = 3.3 $\times$ 1.5 $\simeq$ 5 from the experimental electron band, while thermodynamic studies indicate a value of $\simeq $ 10. It is noted that the possibility of a lower energy interaction within a hundred meV, (for example, the electron-phonon interaction) cannot be ruled out. Unfortunately, the observation of such a small energy scale feature corresponding to the electron-phonon interaction is not possible presently with soft x-ray ARPES from the viewpoint of energy resolution. Nevertheless, the `correlation kink' at 0.25 eV originating in the strongly renormalized bands produce the sharp density of states at the $E_{\rm{F}}$ and indicate the evidence of the mass enhancement in the electronic structure of LaNiO$_3$.  The results indicate a momentum dependent mass renormalization for the electron and hole bands near the $E_{\rm{F}}$, leading to electron-hole asymmetry in this 3-D Fermi liquid.

\section{CONCLUSION}

In conclusion, we have directly observed the 3-D FSs and the energy band dispersion from energy- and angle- dependent PES measurements and determined Re$\Sigma$($\omega$) and Im$\Sigma$($\omega$) of LaNiO$_3$. The energy bands forming the electron FS suggest an $\omega^2$ dependence of the Im$\Sigma$($\omega$) and a `correlation kink' at an energy scale of 0.25 eV. In contrast, the bands which form nesting character hole FSs do not show kinks and match local density approximation calculations. The results indicate a momentum-dependent mass renormalization, leading to electron-hole asymmetry in strongly correlated LaNiO$_3$. Soft x-ray momentum-resolved PES applied to 3-D epitaxial oxide films is expected to provide much needed insights into their electronic structure and exotic properties.





\begin{references}

\bibitem{Imada} M. Imada, A. Fujimori, and Y. Tokura, Rev.\ Mod.\ Phys. {\bf 70}, 1039 (1998).

\bibitem{Shen} A. Damascelli, Z. Hussain, and Z.-X. Shen, Rev.\ Mod.\ Phys. {\bf 75}, 473 (2003).

\bibitem{Ohtomo}A. Ohtomo and H. Y. Hwang, Nature {\bf 427}, 423 (2004).
 
\bibitem{Philip}J. Philip, A. Punnoose, B. I. Kim, K. M. Reddy, S. Layne, J. O. Holmes, B. Satpati, P. R. LeClair, T. S. Santos and J. S. Moodera, Nature Mater. {\bf 5}, 298 (2006).

\bibitem{Ohta}H. Ohta, S-W. Kim, Y. Mune, T. Mizoguchi, K. Nomura, S. Ohta, T. Nomura, Y. Nakanishi, Y. Ikuhara, M. Hirano, H. Hosono, and K. Koumoto,  Nature Mater. {\bf 6}, 129 (2007).

\bibitem{Landau} L. D. Landau, Sov.\ Phys.\ JETP-USSR {\bf 3}, 920 (1957). 

\bibitem{Munoz} J. L. Garc{\'{\i}}a-Mu{\~n}oz, J. Rodr{\'{\i}}guez-Carvajal, P. Lacorre, and J. B. Torrance, Phys. Rev. B {\bf 46}, 4414 (1992).

\bibitem{Zhou} J.-S. Zhou, J. B. Goodenough, B. Dabrowski, P.W. Klamut, and Z. Bukowski, Phys.\ Rev.\ Lett. {\bf 84}, 526 (2000).

\bibitem{LorenzoScagnoli} J. E. Lorenzo, J. L. Hodeau,  L. Paolasini, S. Lefloch, J. A. Alonso, and G. Demazeau, Phys.\ Rev.\ B {\bf 71}, 045128 (2005); V. Scagnoli, U. Staub, A. M. Mulders, M. Janousch, G. I. Meijer, G. Hammerl, J. M. Tonnerre, and N. Stojic,  Phys.\ Rev.\ B\/ {\bf 73}, 100409(R) (2006).

\bibitem{Sreedhar} K. Sreedhar, J. M. Honig, M. Darwin, M. McElfresh, P. M. Shand, J. Xu, B. C. Crooker, and J. Spalek, Phys.\ Rev.\ B {\bf 46}, 6382 (1992).

\bibitem{Xu}X. Q. Xu, J. L. Peng, Z. Y. Li, H. L. Ju, and R. L. Greene, Phys.\ Rev.\ B {\bf 48}, 1112 (1993).

\bibitem{Rajeev} K. P. Rajeev, G. V. Shivashankar, and A. K. Raychaudhuri, Solid State Commun. {\bf 79}, 591 (1991).

\bibitem{ChenDobin} P. Chen, S. Y. Xu, W. Z. Zhou, C. K. Ong, and D. F. Cui, J. Appl.\ Phys. {\bf 85}, 3000 (1999); A. Yu. Dobin, K. R. Nikolaev, I. N. Krivorotov, R. M. Wentzcovitch, E. Dan Dahlberg, and A. M. Goldman, Phys.\ Rev.\ B\ {\bf 68}, 113408 (2003).

\bibitem{Horiba1}K. Horiba, R. Eguchi, M. Taguchi, A. Chainani, A. Kikkawa, Y. Senba, H. Ohashi, and S. Shin, Phys.\ Rev.\ B {\bf 76}, 155104 (2007).

\bibitem{Barman} S. R. Barman, A. Chainani, and D. D. Sarma, Phys.\ Rev.\ B {\bf 49}, 8475 (1994).

\bibitem{HussainChKamakura} Z. Hussain, S. Kono, R. E. Connelly, and C. S. Fadley, Phys.\ Rev.\ Lett. {\bf 44}, 895 (1980); Ch. S{\o}ndergaard, Ph. Hofmann, Ch. Schultz, M. S. Moreno, J. E. Gayone, M. A. Vicente Alvarez, G. Zampieri, S. Lizzit and A. Baraldi, Phys.\ Rev.\ B {\bf 63}, 233102 (2001); N. Kamakura,  Y. Takata, T. Tokushima, Y. Harada, A. Chainani, K. Kobayashi, and S. Shin, Phys.\ Rev.\ B {\bf 74}, 045127 (2006).

\bibitem{VenturiniYanoFujimori} T. Claesson, M. M\r{a}nsson, C. Dallera, F. Venturini, C. De Nada{\"{\i}}, N. B. Brookes, and O. Tjernberg, Phys.\ Rev.\ Lett. {\bf 93}, 136402 (2004);  F. Venturini, J. C. Cezar, C. De Nada{\"{\i}}, P. C. Canfield, and N. B. Brookes, J.\ Phys.\ Condens.\ Matter {\bf 18}, 9221 (2006); M. Yano, A. Sekiyama, H. Fujiwara, T. Saita, S. Imada, T. Muro, Y. Onuki, and S. Suga, Phys.\ Rev.\ Lett. {\bf 98}, 036405 (2007); S. Fujimori, Y. Saitoh, T. Okane, S. Fujimori, H. Yamagami, Y. Haga, E. Yamamoto, and Y. Onuki, Nature Phys. {\bf 3}, 618 (2007). 

\bibitem{NekrasovByczuk}I. A. Nekrasov,  K. Held, G. Keller, D. E. Kondakov, Th. Pruschke, M. Kollar, O. K. Andersen, V. I. Anisimov, and D. Vollhardt, Phys.\ Rev.\ B {\bf 73}, 155112 (2006); K. Byczuk M. Kollar, K. Held, Y.-F. Yang, I. A. Nekrasov, Th. Pruschke, and D. Vollhardt, Nature Physics {\bf 3}, 168 (2007). 

\bibitem{Yoshida} T. Yoshida, K. Tanaka, H. Yagi, A. Ino, H. Eisaki, A. Fujimori, and Z.-X. Shen, Phys.\ Rev.\ Lett. {\bf 95}, 146404 (2005).

\bibitem{Lanzara} A. Lanzara, P. V. Bogdanov, X. J. Zhou, S. A. Kellar, D. L. Feng, E. D. Lu, T. Yoshida, H. Eisaki, A. Fujimori, K. Kishio, J.-I. Shimoyama, T. Nodak, S. Uchidak, Z. Hussain, and Z.-X. Shen, Nature {\bf 412}, 510 (2001).

\bibitem{Valla} T. Valla, T. E. Kidd, W.-G. Yin, G. D. Gu, P. D. Johnson, Z.-H. Pan, and A. V. Fedorov., Phys.\ Rev.\ Lett. {\bf 98}, 167003 (2007).

\bibitem{Meevasana}W. Meevasana,  X. J. Zhou, S. Sahrakorpi, W. S. Lee, W. L. Yang, K. Tanaka, N. Mannella, T. Yoshida, D. H. Lu, Y. L. Chen, R. H. He, Hsin Lin, S. Komiya, Y. Ando, F. Zhou, W. X. Ti, J. W. Xiong, Z. X. Zhao, T. Sasagawa, T. Kakeshita, K. Fujita, S. Uchida, H. Eisaki, A. Fujimori, Z. Hussain, R. S. Markiewicz, A. Bansil, N. Nagaosa, J. Zaanen, T. P. Devereaux, and Z.-X. Shen, Phys.\ Rev.\ B {\bf 75}, 174506 (2007).

\bibitem{Horiba2Ohashi} K. Horiba, N. Kamakura, K. Yamamoto, K. Kobayashi, and S. Shin, J. Electron Spectrosc.\ Relat.\ Phenom. {\bf 144-147}, 1027 (2005); H. Ohashi, Y. Senba, H. Kishimoto, T. Miura, E. Ishiguro, T. Takeuchi, M. Oura, K. Shirasawa, T. Tanaka, M. Takeuchi, K. Takeshita, S. Goto, S. Takahashi, H. Aoyagi, M. Sano, Y. Furukawa, T. Ohata, T. Matsushita, Y. Ishizawa, S. Taniguchi, Y. Asano, Y. Harada, T. Tokushima, K. Horiba, H. Kitamura, T. Ishikawa, and S. Shin,  AIP Proc. {\bf 879}, 523 (2007).

\bibitem{Hamada} N. Hamada, J. Phys.\ Chem.\ Solids {\bf 54}, 1157 (1993).

\bibitem{wien2k} P. Blaha, K. Schwarz, G. Madsen, D. Kvasnicka, and J. Luitz,  WIEN2K, An Augmented Plane Wave+Local Orbitals Program for Calculating Crystal Properties (Karlheinz Schwarz Technical University Wien, Wien, Austria, 2001).

\bibitem{Noun} W. Noun, B. Berini, Y. Dumont, P. R. Dahoo, and N. Keller, J. Appl.\ Phys. {\bf 102}, 063709 (2007).

\bibitem{Munoz2} J. L. Garc{\'{\i}}a-Mu{\~n}oz, M. Suaaidi, M. J. Mart{\'{\i}}nez-Lope and J. A. Alonso, Phys.\ Rev.\ B {\bf 52} 13563 (1995).

\end{references}
\end{document}